\newcommand{\CLASSINPUTbottomtextmargin}{1in}
\newcommand{\CLASSINPUTtoptextmargin}{0.8in}
\newcommand{\CLASSINPUToutersidemargin}{0.63in}

\documentclass[conference]{IEEEtran}
\usepackage{amsmath}
\usepackage[linesnumbered,ruled]{algorithm2e}
\usepackage{graphicx}
\renewcommand{\baselinestretch}{0.99}
\ifCLASSINFOpdf
\else
\fi
\IEEEoverridecommandlockouts
\begin{document}
%
\title{Joint Subcarrier and CPU Time Allocation for Mobile Edge Computing}

\author{\IEEEauthorblockN{Yinghao Yu$^
*$, Jun Zhang$^*$, and Khaled B. Letaief$^{*\dag}$, \emph{Fellow, IEEE}}
\IEEEauthorblockA{$^*$Dept. of ECE, The Hong Kong University of Science and Technology\\
$^\dag$Hamad Bin Khalifa University, Doha, Qatar\\
Email: $^*$\{yyuau, eejzhang, eekhaled\}@ust.hk,$^\dag$kletaief@hbku.edu.qa}
\thanks{This work is supported by the Hong Kong Research Grants Council under Grant No. 16200214.}
}


%


\maketitle


\begin{abstract}
In mobile edge computing systems, mobile devices can offload compute-intensive tasks to a nearby \emph{cloudlet}, so as to save energy and extend battery life. Unlike a fully-fledged cloud, a cloudlet is a small-scale datacenter deployed at a wireless access point, and thus is highly constrained by both radio and compute resources. We show in this paper that separately optimizing the allocation of either compute or radio resource -- as most existing works did -- is highly \emph{suboptimal}: the congestion of compute resource leads to the waste of radio resource, and vice versa. To address this problem, we propose a joint scheduling algorithm that allocates both radio and compute resources coordinately. Specifically, we consider a cloudlet in an Orthogonal Frequency-Division Multiplexing Access (OFDMA) system with multiple mobile devices, where we study subcarrier allocation for task offloading and CPU time allocation for task execution in the cloudlet. Simulation results show that the proposed algorithm significantly outperforms per-resource optimization, accommodating more offloading requests while achieving salient energy saving. 
\end{abstract}


%
\IEEEpeerreviewmaketitle

\section{Introduction}

Limited battery life continuously shows up as the top concern of smartphone users \cite{kumar2010cloud}. The problem is becoming even more severe in the predictable future, given the ever-growing demands for compute-intensive apps and the stalling battery capacity of smartphones. Mobile edge computing recently comes up as a promising solution \cite{barbarossa2014communicating,satyanarayanan2014cloudlets,mao2016dynamic}. By deploying small-scale datacenters at wireless access points -- known as \emph{cloudlets} -- the system allows smartphone users to offload compute-intensive tasks to a nearby cloudlet, so as to extend their battery life by trading off heavy CPU cycles for lightweight communication.

The performance of offloading critically depends on the allocation of both \emph{radio} and \emph{compute} resources: the former determines the data transmission speed and the communication energy consumption; the latter determines the compute time of tasks offloaded to a cloudlet. In general, the more resources are allocated, the better an offloading request is served. However, both radio and compute resources are highly constrained in a cloudlet. In particular, cloudlets are deployed at wireless access points where only a limited number of radio channels are available. Meanwhile, an economic, scalable deployment forces cloudlets to be no more than small-scale datacenters with limited compute capabilities. Therefore, to develop effective computation offloading strategies, it is critical to take both radio and compute resources into account. 


However, a large body of existing works simply assumed an infinite amount of compute resources available in a cloudlet, where the offloaded tasks were computed with negligible processing time. The problem of offloading scheduling was then reduced to radio resource allocation. For example, Chen \emph{et al.} \cite{chen2015decentralized} modeled the competition for radio resources as a congestion game of selfish mobile users. Kaewpuang \emph{et al.} \cite{kaewpuang2013framework} studied the cooperation game of offloading service providers, where the radio and compute resources were assumed to be managed by different entities separately. As we shall show in this paper, coordinately managing both resources improves the overall utilization significantly. Sardellitti \emph{et al.} \cite{sardellitti2015joint}, on the other hand, simply ignored the congestion of compute resources in a cloudlet by throttling the CPU cycles allocated to each offloaded task. Juan \emph{et al.} \cite{liu2016delay} also assumed an infinitely powerful cloudlet such that the execution time for each offloaded task was guaranteed to be a constant value.


Recently, a few researchers started to jointly consider the limitations in radio and compute resources. Nonetheless, some of their assumptions either are inefficient in regards to energy reduction or will weaken the applicability of the result. In \cite{barbarossa2013joint}, CPU resources were allocated as percentages of the total CPU frequency, meaning that jobs are running in parallel. Such parallel execution maintains fairness but prolongs the average execution time. Radio resource was allocated in non-preemptive time slots in \cite{yue2014cloud}. However, all the slots are of a fixed length, and an unnecessarily long slot-length results in waste of radio resource. Furthermore, the efficiency of the proposed scheduling policy is sensitive to some parameters that need to be searched empirically under different system settings. 


Motivated by the above limitations in existing works, in this paper, we propose algorithms that fully utilize the limited radio and compute resources to reduce the energy consumption of mobile devices. Given the small coverage of the cloudlet, we consider an OFDMA system so that interference among users could be ignored, with subcarriers as the radio resource. In terms of compute resource, we allocate CPU time slots of the cloudlet non-preemptively with varied slot-length. We first propose near-optimal algorithms that separately schedule subcarriers and CPU time slots. We show that naive combination of per-resource allocations greatly degrades the system performance. The reason is that congestion of one resource will cause significant waste of the other. To address this problem, we propose a joint scheduling algorithm to coordinately manage subcarriers and CPU of the cloudlet. Simulation results show that a noteworthy amount of energy is saved through joint scheduling compared to separate allocation. Moreover, the coordinate management of different resources is of greater advantage when more prominent performance gains could be achieved through computation offloading.
\begin{figure}[htbp]
\centering\includegraphics[width=5.5cm]{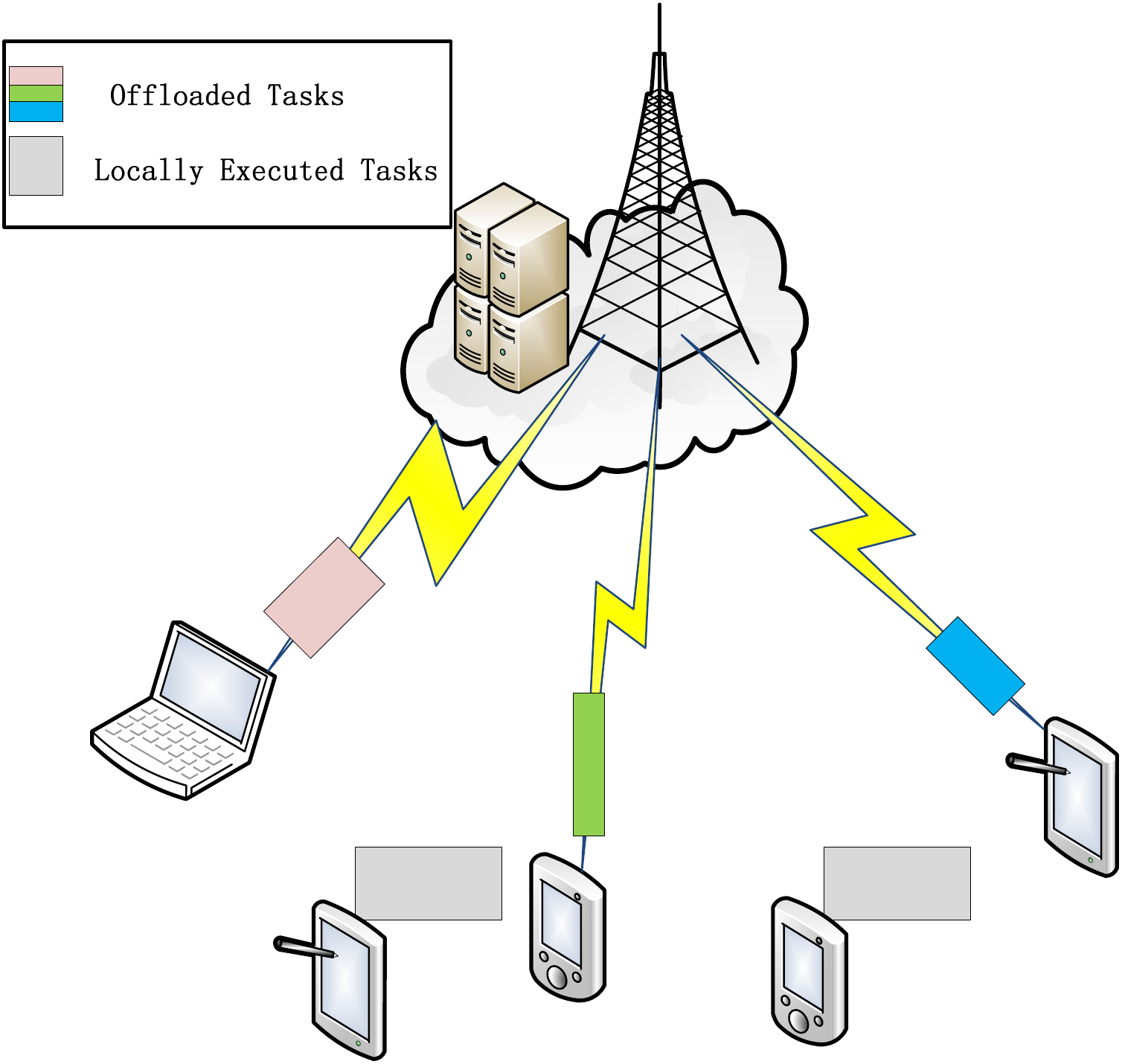}
\caption{A mobile edge computing system with $M$ mobile devices and an infrastructure-based cloudlet.}\label{model}
\end{figure}
\section{System Model}
We consider centralized resource allocation for mobile edge computing with OFDMA as the multiple access scheme. A cloudlet with certain computation capability is deployed at the wireless access point to provide job-execution services. Our objective is to minimize the total energy consumption of mobile users. In this section, we will model both the remote resources in the cloudlet and the local resources of mobile devices. We will then analyze the required energy and time for offloading, and formulate the energy-minimization problem.
\subsection{Model of the Cloudlet and Mobile Users}
As shown in Fig. \ref{model}, we consider a snapshot when the CPU of the cloudlet is idle, and there are $N$ available subcarriers to serve $M$ mobile users. The CPU frequency of the cloudlet is $f_c$. Let $\mathcal{C} = \{1,2,...,N\} $ denote the available subcarriers to be allocated. The bandwidth of each subcarrier is $B_N$. Further denote $\mathcal{G}$ as the channel-gain-to-noise ratio matrix. We assume $\mathcal{G}$ remains constant during the scheduling process. 

Let $\mathcal{U} = \{1,2,...,M\}$ denote the $M$ users, each with a job to execute either locally or remotely. In the following, we may call user and job interchangeably. Each job $\mathcal{J}_i$ is described by its input data size $\mathbf{D}_i$ and deadline $\mathbf{T}_i$. For user $\mathcal{U}_i$, the maximal frequency of local CPU is $\mathbf{F}_i$. Maximal transmission power and static circuit power are denoted by $\mathbf{p}_i^m$ and $\mathbf{p}_i^c$.

\subsection{Energy Consumption}
\subsubsection{Local execution}
According to \cite{zhang2013energy}, at frequency $f_i$, the energy consumption of each CPU cycle is $\kappa f_i^2$, and the required CPU cycles for completing a job is given by $X\mathbf{D}_{i}$, where $\mathbf{D}_i$ is the input data size, while $\kappa$ and $X$ are known constants. In order to minimize the local execution energy consumption, the CPU frequency of $\mathcal{U}_i$ should be set to $f_i=\frac{X\mathbf{D}_i}{\mathbf{T}_i}$ such that its deadline is exactly met since the energy consumption of each CPU cycle increases with its frequency. Thus, the local energy cost for $\mathcal{U}_i$ is given by   
\begin{equation}
E_{l}^i= \kappa f_i^2\cdot X\mathbf{D}_i=\kappa \left(\frac{X\mathbf{D}_i}{\mathbf{T}_i}\right)^2\cdot X\mathbf{D}_i=\kappa X^3\frac{\mathbf{D}_i^3}{\mathbf{T}_i^2}.
\end{equation} 
We assume that $\mathbf{F}_i$ is always larger than $f_i$, so that local execution is always feasible for all users.
\subsubsection{Remote execution}
In the case of offloading, it consumes energy to send the input data $\mathbf{D}_i$ to the cloudlet. The energy consumption for receiving the computation results is ignored as the amount of output data is much less the input data \cite{barbarossa2013joint, yue2014cloud, zhang2013energy}. Therefore, the energy cost for offloading is:
\begin{equation}
E_{r}^i= (\mathbf{p}_i+\mathbf{p}_i^c)\cdot T_{t}^i,
\end{equation} 
where $\mathbf{p}_i$ and $T_{t}^i$ are the transmission power and transmission time, respectively. 

We now show that the optimal value of $\mathbf{p}_i$ could be derived through bisection search. Denote $\mathcal{W}=\{\mathcal{W}(i,j)|\mathcal{W}(i,j)\in \{0,1\},i\in\mathcal{U}, j\in\mathcal{C}\}$ as the subcarrier allocation matrix. For user $\mathcal{U}_i$ who has been allocated a group of subcarriers $\mathcal{W}(i)$, transmit energy efficiency (in bits per joule) is convex with the transmit power \cite{xiong2012energy}. Therefore, via bisection search we can find the optimal $\mathbf{p}_i^*$ that minimizes the transmission energy for the input data. In addition, as the transmit power has to be larger than a threshold $\mathbf{p}_i^t$ to meet the job deadline $\mathbf{T}_i$, we have
\begin{equation}
\mathbf{p}_i= \max(\mathbf{p}_i^*, \mathbf{p}_i^t).
\end{equation}

\subsection{Time for Offloading}
\subsubsection{Transmission}
Let $\mathcal{P}=\{\mathcal{P}(i,j)|\mathcal{P}(i,j)\in [0,\mathbf{p}_i^m], i\in\mathcal{U}, j\in\mathcal{C}\}$ be the power allocation matrix. The optimal power allocation matrix $\mathcal{P}$ is obtained by the water-filling algorithm \cite{wong1999multiuser}. We then have the aggregated data rate as
\begin{equation}
\mathbf{R}_i=B_N\sum_{j=1}^{N}\mathcal{W}_{i,j}\log(1+\mathcal{P}(i,j)\mathcal{G}(i,j)),
\end{equation} 
and transmission time as
\begin{equation}
T_{t}^i= \frac{\mathbf{D}_i}{\mathbf{R}_i}.
\end{equation} 

\subsubsection{Queuing and remote execution}
We assume non-preemptive CPU allocation, which assigns a time slot to one user each time until its job completes. The remote execution time in the cloudlet is then given by
\begin{equation}
T_{c}^i= \frac{X\mathbf{D}_i}{f_c}.
\end{equation} 
 Denote $\mathbf{q}=\{\mathbf{q}_i|\mathbf{q}_i\in \{1, 2, ..., M\}, \mathbf{q}_i \neq \mathbf{q}_j, i, j\in\mathcal{U}\}$ as the execution sequence in the cloudlet, and jobs are executed in the ascending order of $\mathbf{q}$. The queuing time in the cloudlet is then
\begin{equation}
Q_{c}^i =\sum_{j,\mathbf{q}_j<\mathbf{q}_i}^{M}\boldsymbol{\alpha}_j\cdot T_{c}^j,
\end{equation}  
where $\boldsymbol{\alpha}_j$ is the indicator of whether job $\mathcal{J}_j$ is offloaded.
Thus, the total time for remote execution is given by
\begin{equation}
T_{r}^i= T_{t}^i + Q_{c}^i + T_{c}^i.
\end{equation} 
\subsection{Problem Formulation}
We now formulate the total energy consumption minimization problem as follows: 
\begin{align}
&\mathcal{P}: \underset{\boldsymbol{\alpha}, \mathcal{W}, \mathcal{P}, \mathbf{q}}{\mathrm{minimize}} \quad \sum_{i=1}^{M}\left((1-\boldsymbol{\alpha}_i)\cdot E_{l}^i+\boldsymbol{\alpha}_i\cdot E_{t}^i\right),\label{eq1}\\
&\mathrm{subject\thinspace to}  \notag \\
&\sum_{i=1}^{M}\mathcal{W}(i,j)\leq 1,\quad \forall j\in\mathcal{C}   \tag{\ref{eq1}a} \\
&\mathbf{q}_i \neq \mathbf{q}_j, if \  i \neq j. \quad \forall i,\ j \in \mathcal{U} \tag{\ref{eq1}b}\\
&\mathbf{p}_i=\sum_{j=1}^{N}\mathcal{W}(i,j)\mathcal{P}(i,j)\leq \mathbf{p}_i^m,\quad \forall i\in\mathcal{U} \tag{\ref{eq1}c}\\
&\mathbf{R}_i=B_N\sum_{j=1}^{N}\mathcal{W}_{i,j}\log(1+\mathcal{P}(i,j)\mathcal{G}(i,j)),\quad \forall i\in\mathcal{U}\tag{\ref{eq1}d}\\
&E_{l}^i= \kappa X^3\frac{\mathbf{D}_i^3}{\mathbf{T}_i^2} ,\quad \forall i\in\mathcal{U}\tag{\ref{eq1}e} \\
&E_{r}^i=\frac{\mathbf{D}_i(\mathbf{p}_i+\mathbf{p}_i^c)}{\mathbf{R}_i} ,\quad \forall i\in\mathcal{U}\tag{\ref{eq1}f}\\
&T_{r}^i=\frac{\mathbf{D}_i}{\mathbf{R}_i}+ \sum_{j,\mathbf{q}_j<\mathbf{q}_i}^{M}\boldsymbol{\alpha}_jT_{c}^j+\boldsymbol{\alpha}_iT_{c}^i \leq T_i ,\quad \forall i\in\mathcal{U}\tag{\ref{eq1}g}.
\end{align}
Constraint ($\ref{eq1}$a) ensures that each subcarrier is assigned exclusively to one user. ($\ref{eq1}$b) enforces non-preemptive execution in the cloudlet. ($\ref{eq1}$c) and $(\ref{eq1}$d) are the results of bisection search and water-filling with a given the subcarrier allocation matrix $\mathcal{W}$. ($\ref{eq1}$c) places an upper bound for the total transmission power. Finally, ($\ref{eq1}$e) and ($\ref{eq1}$f)  respectively calculate the local-execution and offloading energy and ($\ref{eq1}$g)  enforces the corresponding hard deadline on each of the offloaded task.

This resource allocation problem is a mixed-integer nonlinear programming
(MINLP) problem, which in general is NP-hard. The optimal solution to such a problem is difficult to find, due to the combinatorial optimization variables ($\boldsymbol{\alpha}, \mathbf{q} \text{ and } \mathcal{W} $). Also, handling the non-convex functions in both the objective and constraints brings an additional challenge. In the following, we will propose efficient algorithms to solve this problem with near-optimal performance.
 
\section{Cloudlet with Unlimited Computation Capability}
In this section, we address problem (\ref{eq1}) under a common assumption adopted in existing literatures, i.e., a powerful cloudlet whose computation capability is far beyond the offloading demands of users. For this special case, we develop an efficient algorithm to allocate the radio resources, which could serve as a performance upper bound for the case that the cloudlet possesses limited compute resources as will be pursued in the next section. 

From the previous discussions, the optimal transmit power and power allocation are respectively obtained through bisection search and the water-filling algorithm. The problem now is to allocate the subcarriers properly.

The subcarrier allocation problem in mobile edge computing system poses several new challenges. Firstly, it is difficult to derive closed-form expressions for the outcome of bisection search and water-filling algorithm, which makes it impossible to explicitly compare different subcarrier allocation results. Besides, for the users who execute their tasks locally, the allocated subcarriers will be wasted. 




To avoid such waste of radio resources, subcarriers should be allocated in groups that will ensure beneficial offloading. Intuitively, users with heavy computation workload and meanwhile in good channel conditions should have high priorities to offload, as relatively few subcarriers are required by such users to meet the deadline requirement while energy savings will be large. Thus, to find these users, we propose an algorithm that allocates subcarriers in the ``minimum" group of each user, which is defined as the minimum set of subcarriers required to guarantee beneficial offloading. In each iteration, we find the minimum subcarrier group $\mathbf{C}_i$ for each user $\mathcal{U}_i$ and obtain the corresponding energy consumption. The tasks of the users achieving the most energy savings with their minimum subcarrier groups should be offloaded to the cloudlet. Details of the proposed subcarrier allocation policy are summarized in Algorithm 1. 

\begin{algorithm}
    \SetKwInOut{Input}{Input}
    \SetKwInOut{Output}{Output}

    \Input{$\mathcal{U}=\{1,2,..,M\}, \mathcal{C}=\{1,2,...,N\},\newline \mathcal{G},\mathbf{D},\mathbf{T},\mathbf{p}^m, \mathbf{p}^c,E_l$}
    \Output{$\mathcal{W},\mathcal{P},\boldsymbol{\alpha}$}
	$\boldsymbol{\alpha} \gets \{0,...,0\}.$\\
    \While{$\ |\mathcal{U}|>0\ \mathbf{and}\ |\mathcal{C}|>0$}
      {
	    \For{$i\in \mathcal{U}$}
	 	{  Find the minimum group $\mathbf{C}_i$, such that
 			$E_{t}^i < E_{l}^i, T_{t}^i \leq \mathbf{T}_i,$  where
			$[E_{t}^i, T_{t}^i]=$ subcarrier-Search $(\mathcal{G}(i,\mathbf{C}_i), \mathbf{p}_i^m, \mathbf{p}_i^c,\mathbf{T}_i,\mathbf{D}_i)$
		}
		$m\gets \arg\underset{i}\max\{E_{l}^i-E_{t}^i\};\newline$
		$\mathcal{W}(m)\gets \mathbf{C}_m,\ \mathcal{C}\gets\mathcal{C}-\mathbf{C}_m,\ \mathcal{U}\gets \mathcal{U}-\{m\}\ ;$
		$\boldsymbol{\alpha}_m\gets 1;$
      }
      \For{$j\in\ \mathcal{C}$}
	{
		\For{$i\in\ \mathcal{U}$ and $\boldsymbol{\alpha}_i=1$}
		{
			$[(E_{t}^i)', (T_{t}^i)']=$ 
			Bisection-Search $\newline(\mathcal{G}(i,\mathbf{C}_i\cup\{j\}), \mathbf{p}_i^m, \mathbf{p}_i^c,\mathbf{T}_i,\mathbf{D}_i)$
		}
		$m\gets arg\ \underset{i}\max\{E_{t}^i-(E_{t}^i)'\};\newline$
		$\mathcal{W}(m)\gets \mathbf{C}_m\cup\{j\};$
	}
      return $\mathcal{W}, \mathcal{P}, \boldsymbol{\alpha}$\;
      
    \caption{Minimum-Group Allocation Algorithm}
\end{algorithm}

\section{Cloudlet with Limited Computation Capability}
Though the assumption of unlimited compute resource at the cloudlet simplifies the offloading and subcarrier allocation policy design, it is necessary to consider the limited computation capability of the cloudlets as they are small-scale in practice.  In this case, the queuing delay $Q_{c}^i$ and execution time $T_{c}^i$ in the cloudlet are non-negligible, and the congestion in the cloudlet may lead to the violation of the deadline requirements. Therefore, the compute resources should be properly scheduled to maximize the offloading gain.  In this section, we will first develop a per-resource allocation algorithm that combines the subcarrier allocation policy developed in Section III with an optimal CPU time scheduling strategy, which serves as a baseline. We will then propose a joint scheduling scheme to coordinately allocate subcarriers and CPU time slots.


\subsection{Per-Resource Allocation}
As a baseline, we first consider a per-resource allocation algorithm. In this algorithm, subcarriers are assigned first, and the CPU time slots are scheduled in the second stage. The cloudlet allocates subcarriers following Algorithm $1$, with the only difference on checking deadline constraint, where the execution time in the cloudlet $T_{c}^i$ is also considered. 
Non-preemptive CPU scheduling of the cloudlet in the second stage essentially determines the job execution order. The resulted queuing time of each job dictates whether the deadline requirements are satisfied and finally whether offloading requests are accepted. In the following, we will develop the optimal CPU scheduling policy.

Since non-preemptive CPU scheduling problem is NP-hard \cite{jeffay1991non}, we propose an optimal algorithm based on dynamic programming with pseudo polynomial complexity. In the dynamic programming algorithm, the non-preemptive CPU scheduling problem is decomposed into $M$ states. In state $I$, we solve the subproblem of maximizing total energy saving with $I$ out of the $M$ users, and store the maximum amount of saved energy as well as the corresponding execution time as intermediate results. 

To avoid duplicated iterations, each subproblem adopts previously computed results as input, as elaborated in Algorithm $2$. Assume $\mathbf{U}$ is one of the subsets considered in state $I$. Let $Saving(\mathbf{U})$ and $Time(\mathbf{U})$ be the results of this subproblem. To solve it, we divide all possible execution sequences of these $I$ jobs in $\mathbf{U}$ into $I$ categories by the last executed job. 

\begin{algorithm}
    \caption{CPU Scheduling (Dynamic Programming)}
    \SetKwInOut{Input}{Input}
    \SetKwInOut{Output}{Output}

    \Input{$Saving(\mathbf{U}-\{i\}),Time(\mathbf{U}-\{i\}), for\ i \in \mathbf{U}\newline \mathbf{S}_i, T_c^{i}, \mathbf{T}_i, for\ i \in \mathbf{U}$}
    \Output{$Saving(\mathbf{U}),Time(\mathbf{U})$}

	\For{$i\in\ \mathbf{U}$}
	{
		$tempSaving(i)\gets Saving(\mathbf{U}-\{i\});$
		$tempTime(i)\gets Time((\mathbf{U}-\{i\}));\newline$
		\If{$\max\{T_{t}^i,Time(\mathbf{U}-\{i\})\}+T_{c}^i\leq \mathbf{T}_i$}
		{
			$tempSaving(i)\gets tempSaving(i)+\mathbf{S}_i;$
			$tempTime(i)\gets \max\{T_{t}^i,Time(\mathbf{U}-\{i\})\}+T_{c}^i;\newline$
		}
     }
	$m\gets \arg\underset{i}\max\{tempSaving(i)\};\newline$ 
	$Saving(\mathbf{U})\gets tempSaving(m), \newline Time(\mathbf{U})\gets tempTime(m)$

      
\end{algorithm}

Now consider the category where user $i$ is executed at last. For the first $I-1$ users, the largest energy reduction is $Saving(\mathbf{U}-\{i\})$, and their execution time is $Time(\mathbf{U}-\{i\})$. Note that $Saving(\mathbf{U}-\{i\})$ and $Time(\mathbf{U}-\{i\})$ are collected from the results of state $I-1$. It is unnecessary for all of the first $I-1$ users to offload successfully. For user $i$, the ready time before its job could be executed in the cloudlet is the longer one of queuing time $Time(\mathbf{U}-\{i\})$ and its transmission time $T_t^i$. When the deadline $\mathbf{T}_i$ could be satisfied after the execution in the cloudlet, its offloading request is accepted. The corresponding results of this category will be updated as 
\begin{equation*}
  tempSaving(i) = Saving(\mathbf{U}-\{i\})+\mathbf{S}_i, 
\end{equation*}
and 
\begin{equation*}
tempTime(i)=\max\{T_{t}^i,Time(\mathbf{U}-\{i\})\}+T_{c}^i.
\end{equation*}
Likewise, we calculate the results of all the $I$ categories, and choose the one with the largest energy saving as the final output $Saving(\mathbf{U})$. The results of the final state, where $M$ users are considered, are the optimal solution to the original CPU scheduling problem. 


\subsection{Joint Allocation}
Separate allocation leads to inefficient use of the radio and compute resource because users may still have to execute their tasks locally even if they are assigned with subcarriers. The reason is that in the subcarrier allocation stage, the queuing time for each user remains unknown. The congestion in the CPU of the cloudlet may cause execution deadline violations, leading to the waste of subcarriers. For instance, there might be two users both with stringent deadlines and could not wait for the completion of the other's task. Thus, only the user with a larger energy saving gets the chance of offloading. In this case, subcarriers originally allocated to the user that has to execute its task locally could be re-allocated to other users.

\begin{algorithm}
    \SetKwInOut{Input}{Input}
    \SetKwInOut{Output}{Output}

    \Input{$\mathcal{U}=\{1,2,..,M\}, \mathcal{C}=\{1,2,...,N\},\newline \mathcal{G},\mathcal{D},\mathbf{T}, T_c,E_{l}$}
    \Output{$\mathcal{W},\mathcal{P}, \boldsymbol{\alpha}, \mathbf{q}$}
	$\boldsymbol{\alpha} \gets \{0,...,0\},\ t \gets 0, x\gets1;$\\
    \While{$\ |\mathcal{U}|>0\ \mathbf{and}\ |\mathcal{C}|>0$}
      {

	    \For{$i\in \mathcal{U}$}
	 	{  Find the minimum group $\mathbf{C}_i$, such that
 			$E_{t}^i < E_{l}^i,\ \max\{T_{t}^i,t\}+T_c^i \leq \mathbf{T}_i,$  where
			$[E_{t}^i, T_{t}^i]=\newline$ Bisection-Search $(\mathcal{G}(i,\mathbf{C}_i), \mathbf{p}_i^m, \mathbf{p}_i^c,\mathbf{T}_i,\mathbf{D}_i)$
		}
		$\mathbf{S}_i\gets E_{l}^i-E_{t}^i;$
		$m\gets arg\ \underset{i}\max\{\frac{\mathbf{S}_i}{T_c^i}\};\newline$
		$\mathcal{W}(m)\gets \mathbf{C}_m,\ \mathcal{C}\gets\mathcal{C}-\mathbf{C}_m,\ \mathcal{U}\gets \mathcal{U}-\{m\}\ ;$
		$\boldsymbol{\alpha}_m\gets 1, \mathbf{q}_m\gets x, x\gets x+1, t \gets \max\{T_{t}^i,t\}+T_c^i;$
      }
      \For{$j\in\ \mathcal{C}$}
	{
		\For{$i\in\ \mathcal{U}$ and $\boldsymbol{\alpha}_i=1$}
		{
			$[(E_{t}^i)', (T_{t}^i)']=$ 
			Bisection-Search $(\mathcal{G}(i,\mathbf{C}_i\cup\{j\}), \mathbf{p}_i^m, \mathbf{p}_i^c,\mathbf{T}_i,\mathbf{D}_i)$
		}
		$m\gets \arg\underset{i}\max\{E_{t}^i-(E_{t}^i)'\};$
		$\mathcal{W}(m)\gets \mathbf{C}_m\cup\{j\};$
	}
      return $\mathcal{W}, \mathcal{P}, \boldsymbol{\alpha}, \mathbf{q}$\;
      
    \caption{Joint Allocation Algorithm}
\end{algorithm}

To address the uncertainty of successful offloading and to optimize the utility of limited resources, joint allocation is necessary, i.e., we should find the least amount of both radio and compute resources that ensures successful offloading, and allocate them to the users who save energy most efficiently. 

The proposed algorithm is summarized in Algorithm $3$, where for each user, we find the minimum subcarrier group that supports beneficial offloading and also calculate the amount of CPU time needed for remote processing. The users who save the most energy with each CPU cycle are allocated with the corresponding subcarriers and CPU time slot. 

By identifying the minimum subcarrier group and energy saving per CPU cycle, the utilization of both the radio and compute resources is optimized. Another advantage of joint allocation is that the queuing time is already known before scheduling. Such a prior knowledge of congestion will help to avoid the waste of resources.  

\section{Performance Evaluations}
In this section, we evaluate the performance of proposed algorithms. We focus on four questions: 1) How much energy could offloading save (compared with local execution)?  2) Is joint allocation in advantage of per-resource allocation? 3) How well does the proposed algorithms perform (compared with optimal allocation)? 4) How does the limited computation capability of the cloudlet influence system performance? 
\subsection{System Setting}
Users are randomly located in a circle centered at the cloudlet. Large scale fading of the channels is modeled as: 
\begin{equation*}
PL=20\log(d^{km})+20\log(f^{kHz})+32.45\ (dB).
\end{equation*}
The Rayleigh fading model is adopted for small scale fading. The frequency band of the subcarriers is from $1850$ to $1960$ kHz, with the bandwidth of each subcarrier as $18.75$ kHz. The mobile users' circuit power and maximum transmission power are set to be $50$ mW and $1$ W, respectively. Input data size is uniformly distributed in the range of $900-1100$ bits, the job deadlines are uniformly distributed in the range of $50-150$ ms, and $\kappa$ is set as $1\times10^{-24}$\cite{kumar2010cloud, zhang2013energy}. The parameter $X$ is set to be $18000$ cycles per bit. 
\begin{figure}[tbp]
\centering\includegraphics[width=8.3cm]{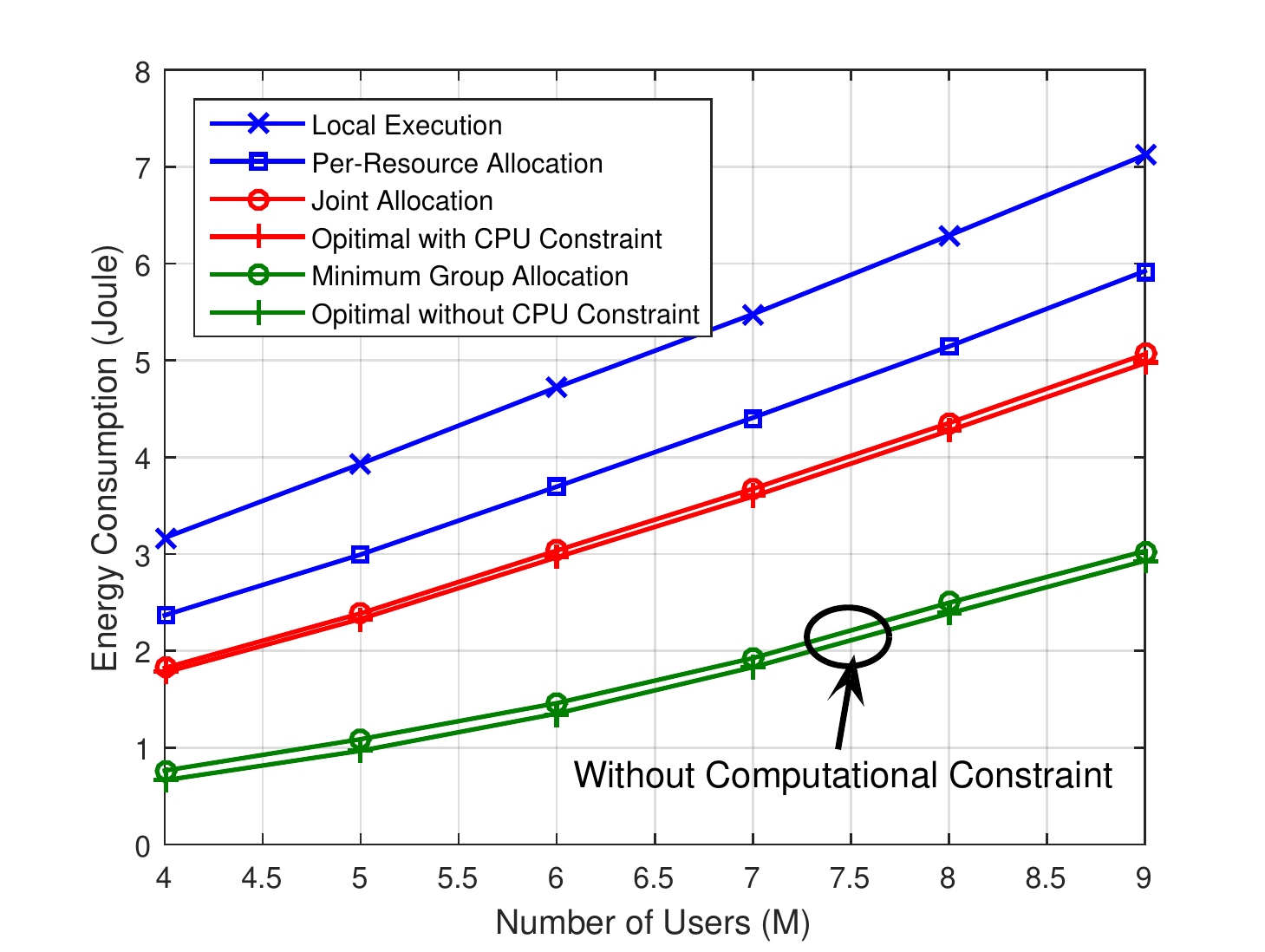}
\caption{Energy Consumption w. r. t. Total Number of Users. $N=4, r=0.2\ \text{km}, f_c=600\ $MHz.}\label{energy_M}
\end{figure}
\begin{figure}[tbp]
\centering\includegraphics[width=8.3cm]{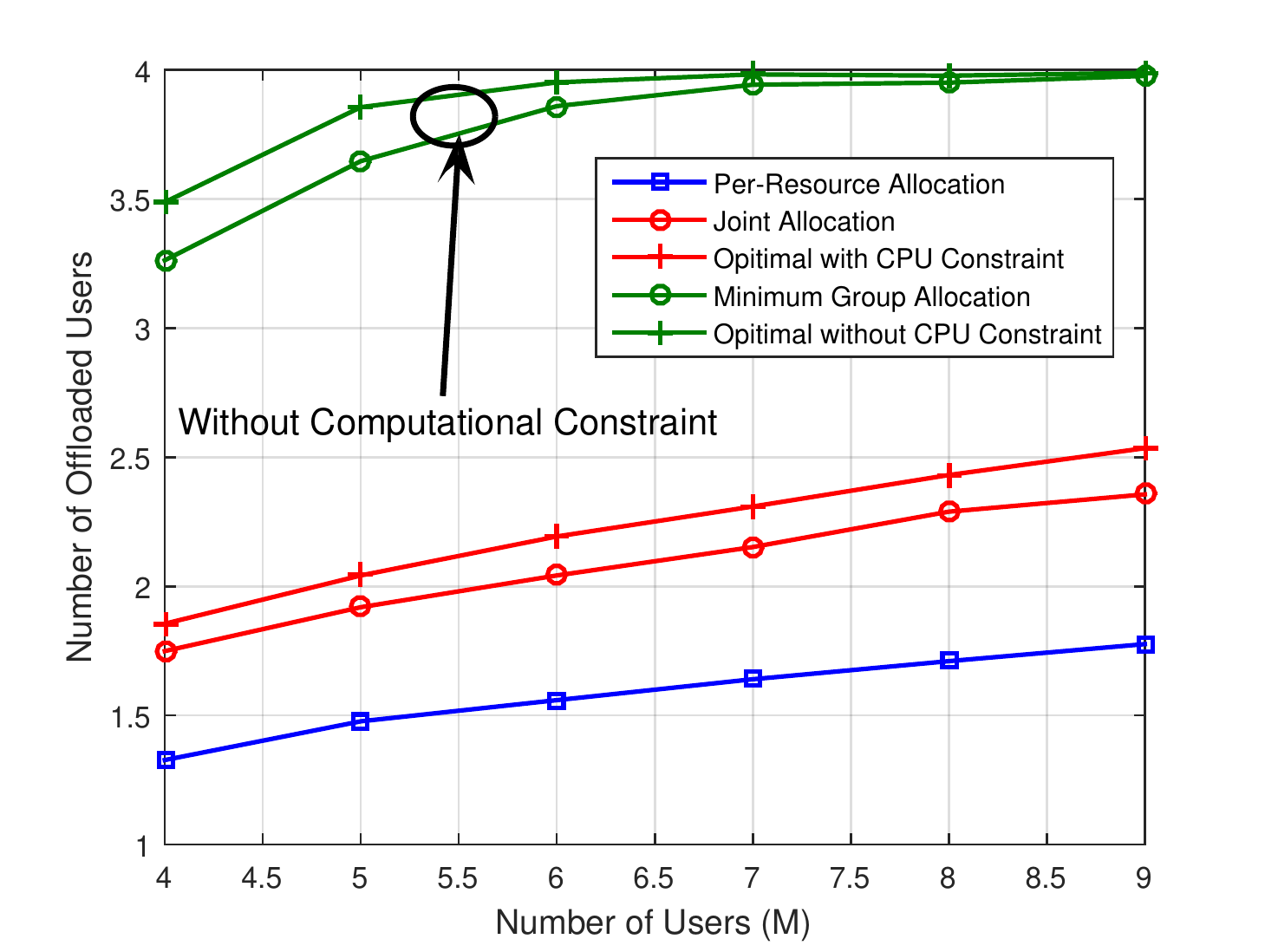}
\caption{Number of Offloaded Users w. r. t. Total Number of Users. $N=4, r=0.2\ \text{km}, f_c=600\ $MHz.}\label{number_M}
\end{figure}
\subsection{Simulations}
In the following simulations, optimal energy savings are obtained by exhaustively searching the subcarrier allocation matrix. In the case when CPU capability is limited, we combine exhaustive search of subcarrier allocation and optimal CPU scheduling to find the optimal results. 
\subsubsection{Number of users}
Energy consumption achieved by the proposed algorithms as the number of users increases are shown in Fig. \ref{energy_M}. From the curves, both Algorithm $1$ (minimum group allocation) and Algorithm $3$ (joint allocation) achieve near-optimal performance, for cases without and with computational constraints, respectively. Per-resource allocation only achieves half of the energy saving compared to joint allocation, although the subcarrier allocation (minimum group allocation) is close to optimal and CPU time allocation (dynamic programming) is optimal. The numbers of offloaded users are compared in Fig. \ref{number_M}, where we find that the constraint of the CPU capabilities (red curves) leads to a reduction of the offloading number by nearly 50\%. 
\subsubsection{Coverage of the cloudlet}
\begin{figure}[htbp]
\centering\includegraphics[width=8.3cm]{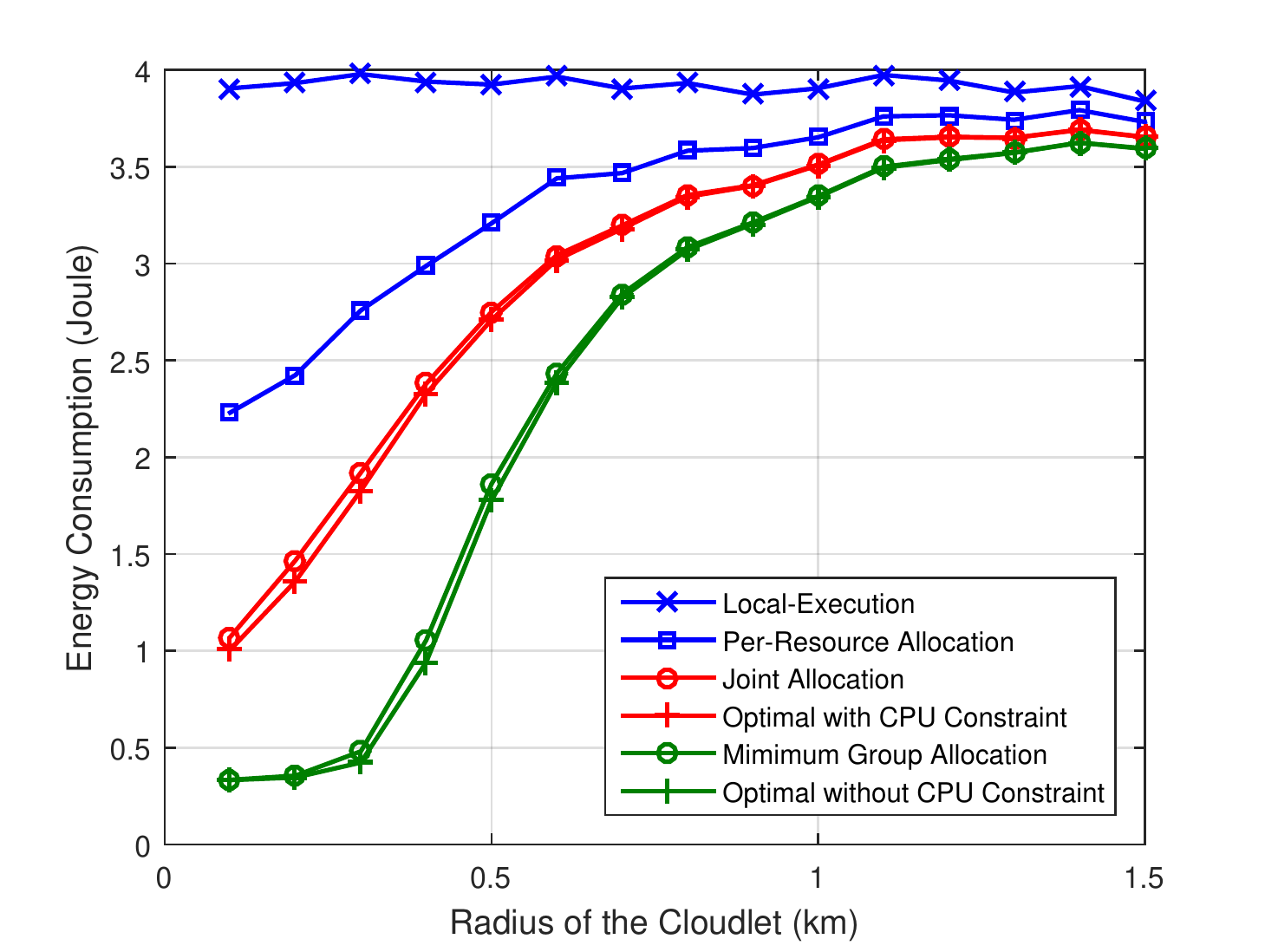}
\caption{Energy Consumption w. r. t. Radius of the Cloudlet. $M=4, N=4, f_c=600\ $MHz.}\label{saving_radius}
\end{figure}

We further investigate the impacts of coverage radius of the cloudlet in Fig. \ref{saving_radius}. As the radius $r$ increases, the offloading gain shrinks. The reason is that the users are distributed at a longer distance from the cloudlet on average. Therefore, fewer users could be supported for offloading. This further demonstrates that to provide satisfactory offloading services and achieve seamless connection, the cloudlets need to be close to the users and densely deployed. Considering the high deployment cost of computationally powerful datacenters, cloudlets with limited compute resource are preferred in practice.

\vspace{-10pt}
\begin{figure}[htbp]
\centering\includegraphics[width=8.3cm]{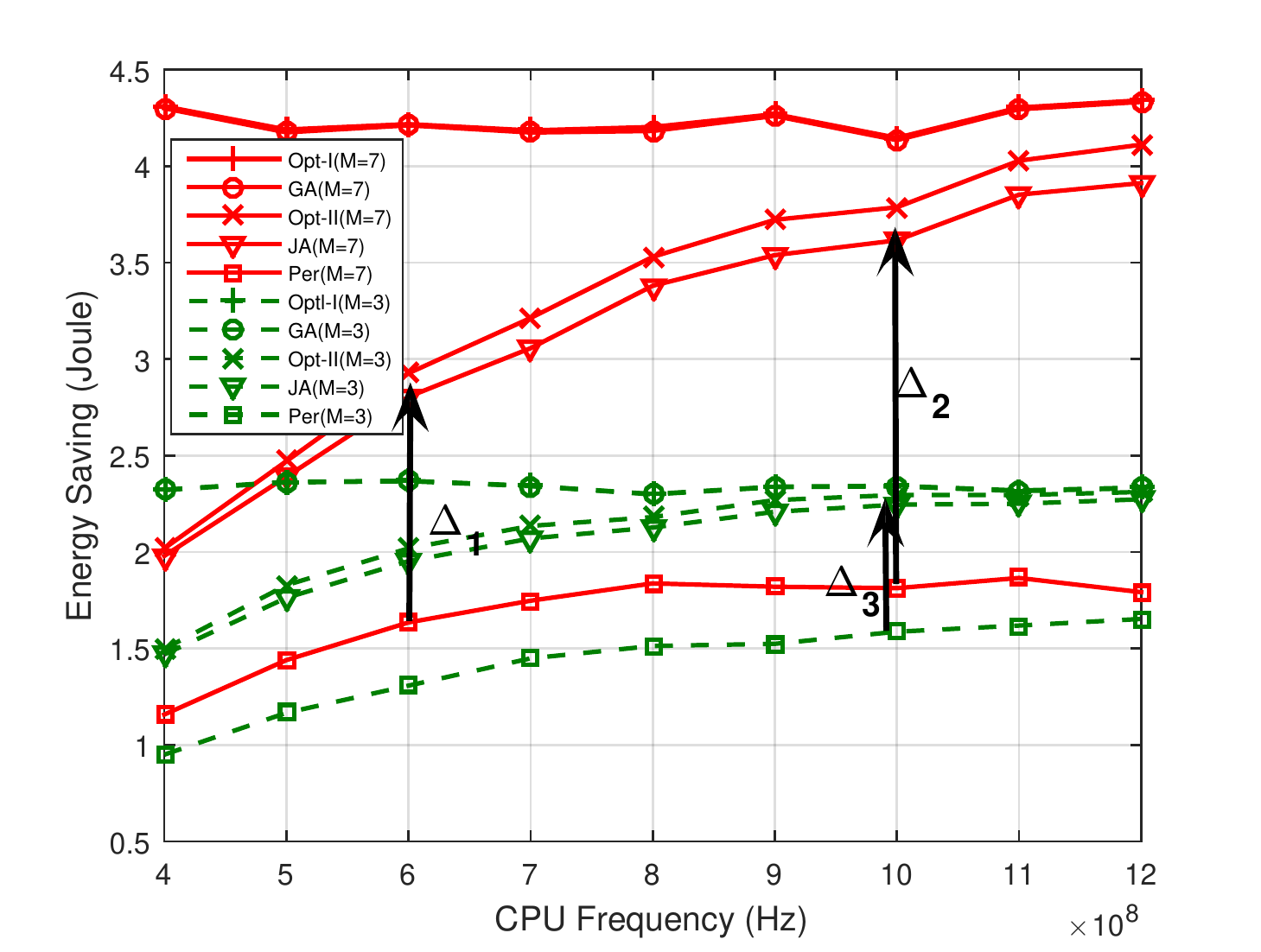}
\caption{Energy Saving w. r. t. CPU Frequency of the Cloudlet. $M=\{3, 7\},\newline N=3$. Legends: Opt-I -- optimal allocation without CPU constraint, Opt-II -- optimal allocation with CPU constraint, MGA -- minimum group allocation, JA -- joint allocation, Per: per-resource allocation.}\label{saving_cpu}
\end{figure}
\subsubsection{CPU frequency of the cloudlet}
Fig. \ref{saving_cpu} shows how the total energy saving varies as computation capability of the cloudlet, i.e., the CPU frequency $f_c$ , increases. It can be seen that there is a saturation point at around $800$ MHz for a 3-user system (green curves). However, when there are 7 users (red curves), the energy saving keeps increasing due to richer user-diversity (comparing $\Delta_2$ with $\Delta_3$). Another observation is that joint allocation is of greater advantage over per-resource allocation as the CPU frequency $f_c$ increases (comparing $\Delta_1$ with $\Delta_2$). When $f_c$ is larger than a threshold, the performance of joint allocation algorithm approaches the optimal scheduling policy without CPU constraint. These results demonstrate that coordinate management better utilizes the resources when a larger offloading gain can be achieved, either by a richer user-diversity or enhanced processing power.

\section{Conclusions}
In this paper, we proposed joint scheduling algorithms of both radio and compute resources for mobile edge computing systems using OFDMA, which are efficient and near-optimal in terms of energy saving for mobile devices. Through extensive simulations, we showed that the per-resource allocation greatly degrades the system performance, even though the allocation policy for each type of resource is near-optimal. Therefore, rather than simply combining separate allocation policies, the congestion information of both types of resources should be considered simultaneously. Furthermore, such joint scheduling is more critical when computation offloading can provide a more prominent energy saving. For future investigations, we will explore the cooperations among cloudlets to further improve the offloading gain. 

\bibliographystyle{IEEEtran}
\bibliography{Edge_Computing}

\end{document}